\documentclass[twocolumn,superscriptaddress,showpacs]{revtex4}
\usepackage{graphicx}

\newcommand{\WMAP}{{\slshape WMAP}}

\begin{document}

\title{Detectability of tensor modes in the presence of foregrounds}

\author{Mihail Amarie}
\email{mamarie@princeton.edu}
\affiliation{Department of Physics, Jadwin Hall, Princeton University, Princeton, New Jersey 08544, USA}

\author{Christopher Hirata}
\affiliation{Department of Physics, Jadwin Hall, Princeton University, Princeton, New Jersey 08544, USA}

\author{Uro${\rm \check{s}}$ Seljak}
\affiliation{Department of Physics, Jadwin Hall, Princeton University, Princeton, New Jersey 08544, USA}
\affiliation{International Center for Theoretical Physics, Strada Costiera 11, 34014 Trieste, Italy}

\date{\today}

\begin{abstract}
In inflationary models
gravitational waves are produced in the early universe and generate $B$-type polarization in the 
cosmic microwave background (CMB). Since $B$ polarization is only generated by gravity waves it does not suffer from the usual cosmic 
variance. A perfect decomposition of the CMB into $B$-modes and $E$-modes would require data from the entire sky, which in practice is 
not possible because of the foreground contaminants. This leads to mixing of $E$ polarization into $B$, which introduces cosmic variance 
contamination of $B$ polarization and reduces
sensitivity to gravity wave amplitude even in absence of detector noise. 
We present numerical results for the uncertainty in the tensor-to-scalar ratio using the Fisher matrix formalism for various 
resolutions and considering several cuts of the sky, using  
the foreground model based on dust maps and assuming 90GHz operating frequency. 
We find
that the usual scaling $\triangle \left( \frac {T}{S} \right) \propto 
f_{sky}^{-1/2}$ is significantly degraded and
becomes $\triangle \left( \frac 
{T}{S} \right) \propto f_{sky}^{-2}$ for $f_{sky}>0.7$. This dependence is affected only weakly
by the choice of sky cuts. 
To put this into a context of what is required level of foreground cleaning, to achieve a $T/S=10^{-3}$ detection at 
3 $\sigma$ one needs to observe 15\% of the sky as opposed to naive expectation of 0.3\%. 
To prevent contamination over this large sky area at required level
one must be able to remove polarized dust emission at or better than 0.1\% of unpolarized intensity, assuming  
the cleanest part of the sky has been chosen. 
To achieve $T/S=10^{-4}$ detection at 3 $\sigma$ one needs to observe 70\% of the sky, which is only 
possible if dust emission 
is removed everywhere over this region at 0.01\% level. Reaching $T/S=10^{-2}$ should be easier: 1\% of the sky is 
needed over which polarized emission needs to be removed at 1\%  of unpolarized intensity if the cleanest 
region is chosen. These results suggest that foreground contamination 
may make it difficult to achieve levels below $T/S=10^{-3}$.

\end{abstract}

\pacs{98.70.Vc, 98.80.Bp, 98.80.Cq}

\maketitle

\section{Introduction}

The polarization in the CMB is generated by Thomson scattering of the CMB photons off free electrons. Thomson scattering generates 
only linear polarization and the polarization can be expressed in terms of the coordinate dependent Stokes parameters $Q$ and $U$. 
It can be also decomposed in the coordinate independent components in the harmonic space: scalar components denoted by $E$, and 
pseudo-scalar components $B$.  For spin $l$, the parity of the $E$-modes is $(-1)^l$, while the parity of the $B$-modes is $-(-1)^l$.  
The primordial density perturbations are scalar only, and because of the parity invariance, they can only generate $E$ modes. The 
gravitational waves present during the inflation are tensor perturbations, so they can generate both $E$ and $B$ modes 
\cite{1997ApJ...482....6S, 1997PhRvD..55.1830Z, 1997PhRvD..55.7368K}. If the amplitude of the gravity waves is very small relative to 
scalars it cannot be isolated from the temperature anisotropies or $E$ polarization due to cosmic variance.  The $B$ polarization is 
insensitive to the cosmic variance from the E modes, being affected only by the foregrounds and the instrument noise. The amplitude of 
the $B$ modes depends on the amplitude of the gravity waves generated during the inflation, which in turn depends on the (as yet 
unknown) energy scale at which the inflation occurred.  The tensor to scalar power ratio is often defined as $T/S \equiv C_2^{TT,\rm 
tensor}/C_2^{TT,\rm scalar}$, i.e. the ratio of the contributions to the CMB temperature quadrupole. In terms of the energy density 
during the inflation it can be written as $T/S \sim V_*/(3.7 \times 10^{16}\,$GeV$)^4$, where $V_*$ is the energy scale of inflation. 
The possibility of detecting direct evidence of gravity waves from inflation via $B$-modes is currently being considered by a number 
of ground, balloon and space based experiments. Notably, a future satellite mission dedicated to $B$ type polarization has been 
identified as one of the NASA Beyond Einstein missions, the Inflation Probe.

There are a number of outstanding issues that need to be resolved for such a mission to become a reality. Some of these are the 
question of required sensitivity and angular resolution to achieve the desired energy scale of inflation below $10^{16}\,$GeV 
\cite{1998PhRvD..57..685K} and the issue of $B$-mode contamination from gravitational lensing \cite{2004PhRvD..69d3005S}. One of the 
main issues related to the feasibility of such observation of the CMB polarization is the contamination by foregrounds. The dust 
present in the interstellar medium (ISM) is one of the main foreground contaminants. It has a thermal vibrational emission spectrum, 
which is increasing with the frequency in the range of interest for us, and it also shows an increased emission in the 10--90 GHz 
range, which could be caused by the rotation of small dust particles \cite{2003NewAR..47.1107L}. The ISM can also be a significant 
source of synchrotron radiation, which shows a red frequency dependence \cite{2003NewAR..47.1117D}. The ISM contaminants are present 
in our galaxy, so the most affected region on a map would be the Galactic Plane.

All these contaminants lead to a more realistic scenario in which we have to exclude portions of the sky because of their 
contamination by foregrounds. This can be done by simply removing some portions of the sky where the contamination is largest, or can 
be more sophisticated by marginalizing over some foreground templates \cite{2004PhRvD..69l3003S, 2004PhRvD..70h3002S}. For simplicity, 
in this paper we only consider the first case. We expect that our conclusion hold qualitatively also for other more sophisticated 
types of analysis and discuss what are required conditions when this is likely to be valid.  
Most CMB parameter forecasting studies have assumed that if only a fraction of the sky $f_{\rm sky}$ is observed 
then the uncertainties in all parameters is reduced by $f_{\rm sky}^{-1/2}$.  However, in case of polarization this ignores an 
important issue. On the whole sky, the $E$ and the $B$ modes are separable. This statement does not hold anymore on a partial sky as 
the boundaries of the cuts generate mixing between these modes. In simple geometries one can isolate pure $E$ modes, pure $B$ modes 
and also mixed modes for which one cannot discern the $E$ or $B$ nature \cite{2002PhRvD..65b3505L, 2003PhRvD..67b3501B}. Pure $E$ 
modes and mixed modes are contaminated by $E$ mode polarization from scalar perturbations, which are larger than the $B$ modes from 
gravity waves even at the present upper bound on $T/S$.  Thus for these modes some of the same cosmic variance issues arise as for $E$ 
polarization and temperature anisotropies. As a result the scaling with $f_{\rm sky}$ can become much worse than expected.  A few 
studies of tensor $B$-modes have taken this issue into account in specific cases such as the \WMAP\ Kp0 cut \cite{2002PhRvD..65b3505L} 
or cuts based on Galactic latitude \cite{2005PhRvD..71f3531H} \cite{Verde:2005ff}, but the full dependence on the sky cut has not been investigated.

Analytic decompositions into $E$, $B$ and mixed modes, emphasized in previous work, are possible only in a few cases of simple 
geometry.  In a more general case with pixelization effects included, all modes are mixed, but the level of mixing varies so that the 
E mode contamination ranges from negligible to nearly complete. Qualitatively, the larger the area of the sky coverage and the smaller 
the length of the boundaries, the smaller the contamination and mode mixing, but analytic results are difficult to obtain. Fortunately 
the decomposition into pure and mixed states, while useful for heuristic interpretation, is not necessary in the actual analysis of 
the data. One simply needs to perform the usual likelihood analysis of the data expressed with measured Stokes $Q$ and $U$ parameters. 
The likelihood analysis is an optimal analysis and as such cannot be improved upon by performing the decomposition into pure and mixed 
modes.

Broadly speaking there are two contributions to $B$-mode polarization. One is from the recombination epoch (which defines the last 
scattering surface), which leads to a peak in the power spectrum at $l \sim 100$.  The second is from the reionization, which 
re-scatters the CMB quadrupole on the horizon scale during reionization and leads to a large signal at large angular scales, $l<20$, 
while at the same time suppressing the recombination signal at smaller scales\cite{1997PhRvD..55.1822Z}. If the optical depth to 
reionization is high, as suggested by \WMAP\ first year results \cite{2003ApJS..148....1B, 2003ApJS..148..161K, 2003ApJS..148..175S, 
2005PhRvD..71j3515S}, then the latter signal dominates in the determination of the gravity wave amplitude \cite{1998PhRvD..57..685K}.  
Because of its larger angular scale, this component of the signal also suffers more from $E$-mode mixing into $B$-modes due to 
incomplete sky coverage. In this paper we include both, varying the amplitude of optical depth over the allowed range.

\section{Quadratic Estimator and Fisher matrix}

We consider the quadratic estimator for the vector $x=(Q_1,Q_2,...Q_N,U_1,U_2,...U_N)$, where N is the total number of pixels 
considered and Q and U are the coordinate dependent Stokes parameters. The covariance matrix $C=\langle x x^{\dagger} \rangle$ is 
written as
\begin{equation}
C_{ij} = C_{0,ij} + \frac{T}{S} C_{T/S,ij}.
\end{equation}
Here ${\bf C}_0$ is the covariance matrix including noise and scalar fluctuations but no tensors, $T/S$ is the tensor-to-scalar ratio, 
and ${\bf C}_{T/S}$ is the $B$-mode power spectrum template, 
\begin{equation}
C_{T/S,ij} = \sum_{lm} \frac{\partial C_l^{BB}}{\partial (T/S)}
{\bf Y}^{B\,\ast}_{lm}({\bf n}_i){\bf Y}^B_{lm}({\bf n}_j).
\label{cts}
\end{equation}
We have left out the $E$-mode power spectrum from the tensors, which contributes negligibly at small $T/S$ due to cosmic variance.
In our case, the ${\bf C}_0$ matrix contains both the contribution from the detector noise and the $E$-mode contribution to the polarization. We work in the limit of $T/S=0$, so we do not consider the second term 
for the covariance matrix. The quadratic estimator determines the tensor-to-scalar ratio using the relation
\begin{eqnarray}\label{FisherMatrix}
\nonumber \widehat{T/S}&=&[ F^{-1}]q \\
\nonumber q &=&\frac{1}{2}{\bf x}^{\dagger}{\bf C}_0^{-1}{\bf C}_{T/S}
{\bf C}_0^{-1}{\bf x}-\frac{1}{2}{\rm Tr}\left({\bf C}_0^{-1}{\bf C}_{T/S}\right) \\ 
F&=&\frac{1}{2}Tr \left({\bf C}_0^{-1}{\bf C}_{T/S}{\bf C}_0^{-1}{\bf C}_{T/S}\right);
\end{eqnarray}
here $F$ is the $1\times 1$ Fisher matrix and ${\bf C}_0$ can be any positive definite hermitian matrix. In our case the Fisher matrix is 
reduced to only 
1 element, since we have only 1 parameter (T/S). For any choice of ${\bf C}_0$ the estimator will be unbiased, however the quadratic 
estimator method has optimum efficiency if ${\bf C}_0$ is the covariance matrix itself. In our particular case ${\bf C}_0$ contains 
the instrument noise which is uncorrelated and also contains the contribution of the scalar modes to the power spectrum. As we are 
interested in estimating $\triangle \left( \frac{T}{S} \right)$, we want to evaluate the Fisher matrix element $F$. In this case 
${\bf C}_{T/S}$ contains just the $B$-modes contribution to the power spectrum. Equation (\ref{FisherMatrix}) requires the signal and 
the noise to be Gaussian. 
The gravitational wave signal is Gaussian because it comes from the fluctuations of a linear quantum oscillator.  Instrument noise is 
typically Gaussian due to the central limit theorem; it is assumed here for simplicity that it is white but this assumption may be 
inappropriate for some experiments.

\begin{figure}
\includegraphics[width=3.2in]{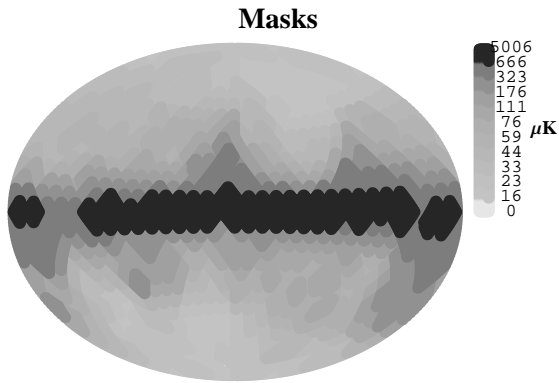}
\caption{\label{masks}
The masks used for cutting the sky based on HEALPix resolution 3 galactic dust maps in Aitoff projection. Various levels of 
gray show the way the masks were built by adding areas with higher dust temperature. The units are of black body temperature.}
\end{figure}

The Fisher matrix error $\triangle \left( \frac{T}{S} \right)$ is the 1$\sigma$ standard deviation of $\widehat {T/S}$ assuming that 
$T/S=0$. The caveats of this method are that the error distribution of $\triangle \left( \frac{T}{S} \right)$ is non-Gaussian and it 
depends on $T/S$ (in particular if $T/S$ is increased, the error goes up).  During our calculations, we used the galactic dust 
temperature map \cite{1999ApJ...524..867F} to create the masks for our patches. For all the resolutions 
considered in our calculations, we used the same dust map, sampled at a lower resolution. As we included a higher portion of the sky in 
our calculations, we kept adding regions contaminated more and more by the foregrounds in our map. They are based on masks, some of 
which are shown in Fig.\ref{masks}. For one of the calculations, in order to point out the effect of having a shorter length for the 
boundary of the cut, we used cuts parallel to the galactic plane.

\section{Numerical Method}

\begin{figure}
\includegraphics[width=3.2in]{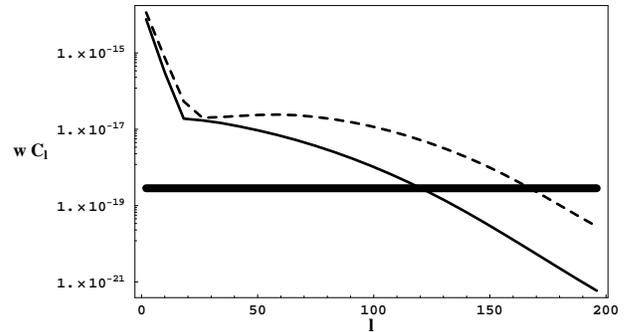}
\caption{\label{cls}
${\rm w}C_l^{EE}$, ${\rm w}C_l^{BB}$ and the lensing + detector noise level after applying the Gaussian window for $T/S \simeq 1/2$.
${\rm w} C_l^{EE}$ is represented by the dashed line, ${\rm w} C_l^{BB}$ by the continuous line and the lensing + detector noise level
by the thick horizontal line. These are the values used for resolution 6, where $l_{Ny} = 196$.}
\end{figure}

We have developed two different numerical methods for evaluating Eq. \ref{FisherMatrix}, both of them having as input the scalar and 
tensor modes generated by CMBFAST \cite{1996ApJ...469..437S} with the RECFAST recombination routines \cite{1999ApJ...523L...1S, 
2000ApJS..128..407S}. We are using maps of various resolutions in the HEALPix \cite{1999elss.conf...37G, 2005ApJ...622..759G} 
pixelization. Given the resolution $r$, the number of pixels a HEALPix map has is $N=12(4^r)$.  The pixel area is
$A=4\pi/N$, and the side of the pixel is
\begin{equation}
d=\sqrt{A}=\sqrt{\frac{4 \pi}{N}}.
\end{equation}
The Nyquist mode will be
\begin{equation}
l_{Ny}=\frac{1}{2}\frac{2 \pi}{d}=\frac{\sqrt{\pi N}}{2}
\end{equation}
We are using both the scalar and the tensor modes with $2 \leq l \leq l_{Ny}$. $l_{Ny}$, and we apply a Gaussian window on them of 
the form
\begin{equation}\label{alfa}
{\rm w}=\exp \left[ \frac{- \pi l (l+1)}{3 \alpha N} \right];
\end{equation}
here $\alpha$ is a Gaussian window factor. To avoid high-$l$ $E$-modes leaking down to low $l$ through pixelization aliasing effects, 
we require that ${\rm w}C^{EE}_{l_{Ny}}\ll C^{noise}$, which setting $C^{noise}/({\rm w}C^{EE}_{l_{Ny}})=10$ gives
\begin{equation}
\alpha = \frac{\pi^2}{12 \left( \ln \frac{C^{EE}_{l_{Ny}}}{C^{noise}}+\ln \left( 10\right) \right) };
\end{equation}
and in Fig.\ref{cls} we can see the window function satisfying the above requirements applied to the $E$ and $B$-modes.
\begin{figure}
\includegraphics[width=3.2in]{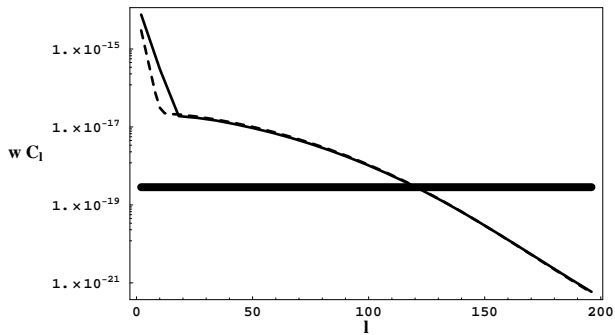}
\caption{\label{reion}
${\rm w}C_l^{BB}$ for two different values of the optical depth to the reionization $\tau$. The dashed line corresponds to a value 
of $\tau = 0.07$ and the continuous line to $\tau = 0.17$. The thick horizontal line is the lensing + detector noise level.}
\end{figure}

The exact method uses the convolution algorithm of Ref.~\cite{2004PhRvD..70j3501H}
to generate the ${\bf C}_0$ matrix starting from $C^{EE}_l$, $C^{noise}$ and to implement Eq.~(\ref{cts}) for the
${\bf C}_{T/S}$ matrix from $C^{BB}_l$.  After generating the matrices, we want to compute $\frac{1}{2}{\rm Tr}\left[({\bf C}_0^{-1} 
{\bf C}_{T/S})^2 \right] $. We use the Cholesky 
method for computing the inverse of ${\bf C}_0$ and then the back substitution method for the multiplication with ${\bf C}_{T/S}$. The 
final step 
is taking the square of the final matrix, which is done classically. As we want to evaluate the same quantities for different cuts of 
the sky, we can save processor time, by doing the computation progressively increasing the area which is covered. The computer memory 
is another limiting factor for our computation. Given that $N$ is the number of pixels, this method needs to store in the memory 
$3 N^2/4$ double precision numbers, to the leading order of $N$. With 4GB of memory available, $r = 5$ is the highest  
resolution at which we can perform the exact calculation. The exact method is an $N^3$ process, which makes the processor time 
required for higher resolutions to increase rapidly. 

In addition to the method described above, we are using a Monte Carlo method to evaluate the Fisher matrix. 
It evaluates
\begin{equation}
F_{T/S,T/S} = \frac{1}{2}\langle {\bf x}^{\dagger}{\bf C}_0^{-1}{\bf C}_{T/S}{\bf C}_0^{-1}{\bf C}_{T/S}{\bf x}\rangle,
\end{equation}
where {\bf x} is a randomly generated vector and ${\bf C}_0^{-1}{\bf x}$ is computed iteratively, without having to take the inverse of the ${\bf C}_0$ 
matrix. We evaluate ${\bf C}_0^{-1}{\bf x}$ using the un-preconditioned conjugate gradient method with the spherical harmonic 
transform routines described in Ref.~\cite{2004PhRvD..70j3501H}.
We compute the mean for different realizations of ${\bf x}$ and use the standard deviation to obtain $1\sigma$ errors.
This method is more efficient than the exact approach
for resolutions higher than 4 as it scales as $\propto N^{3/2}$ (the number of conjugate-gradient iterations needed to solve 
${\bf C}_0^{-1}{\bf x}$ and the the number of realizations of ${\bf x}$ needed to compute the trace are roughly 400 and 25, 
respectively, and do not depend significantly on the resolution, but depend more on the number of pixels chosen within the mask and 
also on the choice of the mask).
We chose a value of $5.0\,\mu$K arcmin for the noise, which is the contribution of the lensing noise level. The 
results can be scaled down to lower lensing + detector noise if lens cleaning is used \cite{2002PhRvL..89a1303K, 
2002PhRvL..89a1304K, 2004PhRvD..69d3005S}, e.g. a $2.5\,\mu$K arcmin experiment would have an uncertainty $\triangle(T/S)$ 
that is reduced by a factor of 4.

\section{Results}

We have computed the Fisher matrix using Eq.~(\ref{FisherMatrix}), and from this, we can calculate
\begin{equation} \label{Delta}
\triangle \left( \frac{T}{S} \right) = F_{T/S,T/S}^{-1/2};
\end{equation}
We used both methods for resolutions 4 and 5, while for resolution 6, we used only the Monte Carlo method. In order to measure more
accurately the tensor-to-scalar ratio, a scientific mission would need to cover a higher fraction of the sky. We make plots showing
the uncertainty in the tensor-to-scalar ratio ($\triangle \left( \frac{T}{S}\right)$) versus the sky fraction we need to cover in
order to achieve this uncertainty. As the Monte Carlo method does not output the exact value of the Fisher matrix, but it only
estimates it, every time we used the Monte Carlo method, the points are presented with the corresponding 1$\sigma$ error bars. The
area of the sky used in our calculations is chosen based on the masks which were presented in a previous section (Fig.\ref{masks}).

\begin{figure}[!h]
\includegraphics[width=3.2in]{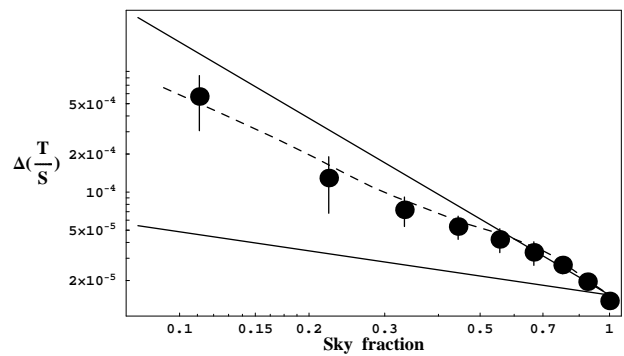}
\caption{\label{res4}
$\triangle \left( \frac{T}{S} \right)$ as a function of sky fraction. The dashed line is computed based on the exact 
method for resolution 4, while the points with error bars are  based on the Monte Carlo method for resolution 6 The continuous lines 
represent the theoretical scaling of $\triangle \left( \frac{T}{S} \right)$. The lower line corresponds to the idealized 
scaling $\triangle \left( \frac{T}{S} \right) \propto f_{sky}^{-1/2}$, while the upper line
is $\triangle \left( \frac{T}{S} \right) \propto f_{sky}^{-2}$ and fits better the actual results for $f_{sky}>0.7$.}
\end{figure}

For the entire sky, the uncertainty in the tensor-to-scalar ratio is about $1.5 \times 10^{-5}$. According to the usual scaling 
$\triangle \left( \frac{T}{S}\right) \propto f_{sky}^{-1/2}$, for a 70\% coverage of the sky we should obtain $\triangle \left( 
\frac{T}{S}\right) = 1.8 \times 10^{-5}$; instead, from our calculations we get $\triangle \left( \frac{T}{S} \right) = 3.2 \times 10 
^{-5}$, nearly a factor of 2 degradation over the idealized case. 
For $f_{sky}>0.7$ the scaling is $f_{sky}^{-2}$, as it is depicted in Fig.\ref{res4} by the straight diagonal line. While the scaling is less steep below that we still find it is steeper 
than expected. To achieve a 3 $\sigma$ detection of $T/S=10^{-3}$ we need 15\% sky coverage, compared to 0.3\% from 
idealized scaling, while for $T/S=10^{-2}$ the required sky coverage is 1\%. 

To interpret these results the most relevant question is what constraints do these results place on the required 
level of foreground subtraction. Typically in CMB we remove some part of the sky where foregrounds are just too
difficult to deal with, while in the remaining area some foreground subtraction method is applied to reduce the 
contamination below the required level. This level of course depends not only on the foregrounds, but also 
on the expected signal. 
To get a lower $\triangle \left( \frac{T}{S} \right)$ we have to map a larger fraction of the sky, but this means that we have to get 
into more contaminated regions of the sky, which requires better methods for cleaning the foreground. 
The cleaning methods are based on expected frequency scaling of dust radiation, which differs from that of CMB. However, 
frequency scaling may not be the same at every direction because of a different composition of dust 
grains. This leads to a decorrelation of dust maps at differing
frequencies. This provides a fundamental limitation to foreground cleaning that is very difficult to circumvent. 
Since
we do not have data about the polarization of dust radiation we will just assume different levels for it.
We also do not know how decorrelated dust maps are as a function of frequency separation. Rough expectation is that 
dust is polarized at a few percent level, so we expect that cleaning at 1\% should be 
relatively straightforward, at 0.1\% significantly more challenging and at 0.01\% may be impossible.
The tensor-to-scalar ratio for which the polarized signal is equal to the polarized fraction of the dust thermal emission is defined as:
\begin{equation} \label{bdef}
\frac{T}{S}=\frac{\left(p T_{dust}\right)^2}{\sum_{l=2}^{l_{max}} \frac{2l+1}{4 \pi}C_l^{BB}|_{T/S=1}};
\label{b}
\end{equation}
where $p$ is the polarized fraction of the dust thermal emission.
We will assume we work at 90 GHz, where dust signal is relatively low and we ignore the
contribution from polarized synchrotron emission, which is expected to be similar to or slightly less than the dust polarization at 
90 GHz \cite{2005MNRAS.360L..10C}.  Because of this many of the upcoming polarization 
experiments are expected to operate at higher frequencies where dust signal is stronger relative to CMB, 
in which case our results may be overly optimistic and need to be rescaled appropriately.  
Note that our analysis is simplistic in the sense that we only use Eq.~(\ref{b}) to asses the 
detectability prospects and do not include a more detailed information on how the two signals 
vary with scale. It should nevertheless give a good ballpark estimate of the required levels of decontamination. 

\begin{figure}
\includegraphics[width=3.2in]{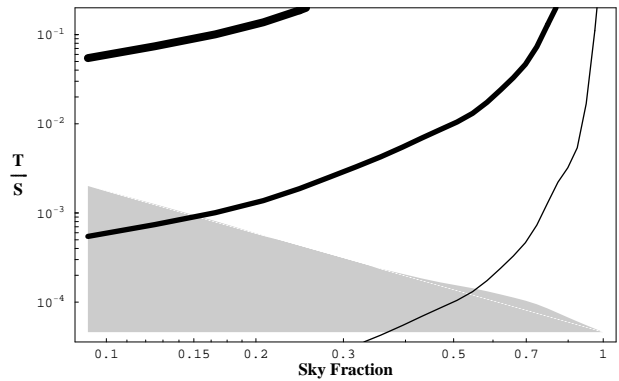}
\caption{\label{blines}
The lines represent the fraction of the sky for which the polarized fraction of the dust thermal emission is less than the polarized 
signal, as defined in equation \ref{b}, assuming the polarized fraction of the dust thermal emission to be 1\% (top), 
0.1\% (middle) and 0.01\% (bottom).
Dashed area is excluded if we require a detection of tensor-to-scalar ratio $\frac{T}{S}$ at a $3\sigma$ confidence. 
For example, to detect $T/S=10^{-3}$ at $3\sigma$ 
we need to observe 15\% of the sky and if that is chosen to be the cleanest region of the sky then 
the dust polarization in this region needs to be cleaned at 0.1\% level of its intensity.} 
\end{figure}
For example, if we want to detect a tensor-to-scalar ratio of $10^{-4}$, we would need to map 70\% of the sky and the polarized 
fraction of the dust thermal emission needs to be less than 0.01\%., 
while if we want to detect a tensor-to-scalar ratio of $10^{-3}$, we 
need to map 15\% of the sky and the polarized fraction of the uncleaned 
dust thermal emission in this region cannot exceed 0.1\% (Fig.\ref{blines}).

\begin{figure}
\includegraphics[width=3.2in]{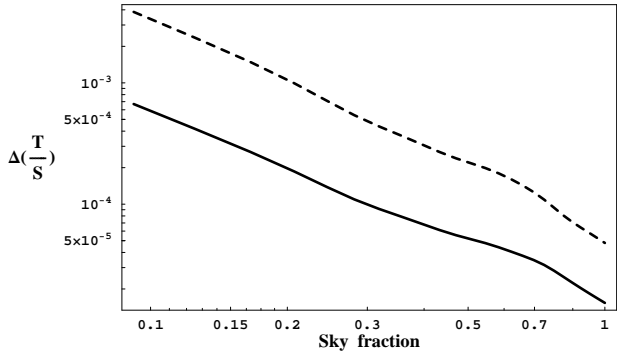}
\caption{\label{res7}
$\triangle \left( \frac{T}{S} \right)$ computed based on the exact method for 
resolution 4. The dashed line corresponds to an optical depth $\tau = 0.07$, while the continuous line corresponds to $\tau=0.17$.} 
\end{figure}

Besides parameters that depend on the instrument, there are also cosmological 
parameters that can influence the detection of $B$-modes. As it was discussed in the Introduction section of the paper, a 
cosmological parameter which has a large influence on the results of our calculations is the optical depth to the reionization, 
$\tau$. For example, if we vary the value of $\tau$ from 0.17, to 0.07, for HEALPix resolution 4 the value of $\triangle \left( 
\frac{T}{S} \right)$ increases by half an order of magnitude,wile the effect of varying the optical depth on the $B$-modes can be seen in Fig.\ref{reion}. We show $\triangle \left( \frac{T}{S} \right)$ for different values of 
optical depth $\tau$ in Fig.\ref{res7}.

\begin{figure}
\includegraphics[width=3.2in]{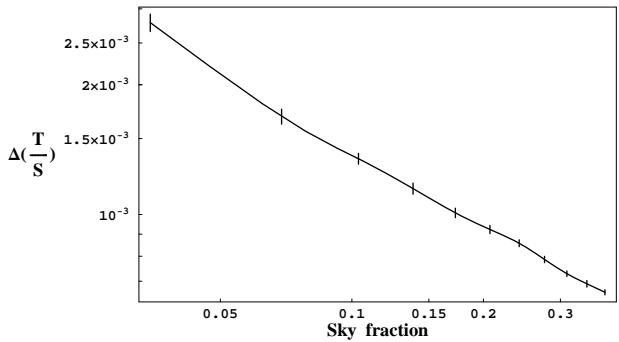}
\caption{\label{ground}
$\triangle \left( \frac{T}{S} \right)$ for HEALPix resolution 7 for a map confined to Earth's southern hemisphere, declinations 
$\delta<-30^{\circ}$. The graph is obtained by ignoring the reionization peak for the $B$-modes and using the Monte Carlo method, 
with $1\sigma$ error bars shown on the graph. The continuous line is a cubic spline through the considered points.}
\end{figure}

As space missions require a large budget, the scientific community is also considering ground-based or balloon-borne experiments. They
can achieve a higher angular resolution, but their sky coverage has to suffer. We are also considering this case, by doing a
calculation at HEALPix resolution 7, confined to declinations $\delta<-30^{\circ}$ around the terrestrial south pole. For the purpose
of this calculation, we ignored the contribution of the reionizatoin peak to the tensor modes, by considering only $C_l^{BB}$ with
$l>20$. The results of this calculation can be seen in Fig.\ref{ground}. Thus, in order to be able to detect $\triangle
\left( \frac{T}{S}\right)=10^{-2}$ at $3\sigma$, we need to look at 1\% of the sky and from Eq.\ref{bdef} we find that the polarized dust signal must be cleaned down to 1\%  of its intensity.

\section{Discussion and conclusions}

\begin{figure}[!h]
\includegraphics[width=3.2in]{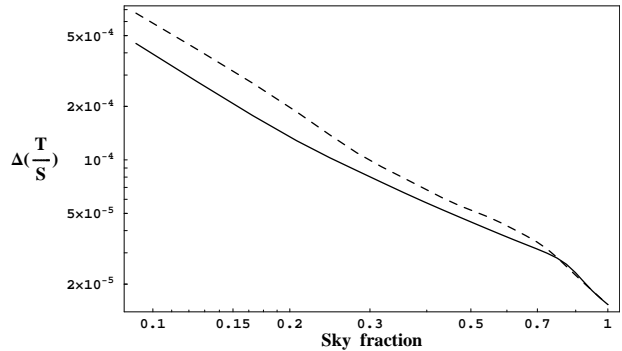}
\caption{\label{cut}
The difference in $\triangle \left( \frac{T}{S} \right)$ for two different types of cuts. The continuous line is obtained 
for the parallel equatorial cut. Both results are obtained through the exact method for HEALPix resolution 4.}
\end{figure}

The main conclusion we get from these results is that the Fisher matrix element $F_{T/S,T/S}$ does not depend linearly on 
the sky percentage considered in the map. Instead, it depends more like $F_{T/S,T/S} \propto f_{sky}^4$. This is in some way expected as the 
harmonic decomposition of the sky is nonlocal and cutting off some parts of the sky affects the rest of it also especially for the 
large scale where the $B$-modes are more important. It is not an exact power law dependence, but as it can be seen from Fig.\ref{res4}, the power law could be a good approximation. We fitted the points from 10\% to 100\% in sky fraction to a power law. There is a disagreement between this dependence and the one from FIG.8 of Ref. \cite{2002PhRvD..65b3505L}, where $F_{T/S,T/S} \propto f_{sky}^{1.7}$. The method proposed in Ref. \cite{2002PhRvD..65b3505L} uses projected $E$ and $B$-modes, which triggers loss of information in the low multipoles. In the pixel space, this loss increases for small sky patches. To give an idea of the way low multipoles affect our analysis, in the case where the effect of the $B$-modes in the reionization peak was not considered (Fig.\ref{ground}), $F_{T/S,T/S} \propto f_{sky}^{1.2}$. 
We considered different pixelizations by changing the resolution of our HEALPix maps and we can see from Fig.\ref{res4} that this does not affect much the noise levels, leaving the results unchanged. The way we choose to make the cut is mildly important in evaluating the results. If instead of 
using a cut based on the galaxy dust maps, we take a straight symmetric cut, parallel to the galactic equator, we get higher 
$F_{T/S,T/S}$ values, but the change is modest (Fig.\ref{cut}).

Our results suggest that to achieve a detection of $T/S$ at $10^{-3}$ level one needs to observe 15\% of the whole sky as opposed to 
0.3\% naively expected. This makes the case for a space mission stronger if such levels of $T/S$ can be motivated theoretically. 
However, this is somewhat dependent on the assumed optical depth, which we do not know yet with enough precision.
Moreover, it is unclear if we can actually foreground clean the observed portion of the sky sufficiently well; as we have 
indicated the required levels are challenging indeed, in the case of $T/S=10^{-3}$ they require a factor of 10-100
reduction of polarized dust emission even if the cleanest part of the sky is chosen, agreeing with the results from \cite{Verde:2005ff}, where the authors considered several specific situations.
It is premature to give a precise verdict in favor of a space mission or a ground-based /balloon-borne experiment, what is 
clear is that we need more data from the experiments currently running in order to put a 
better constraint on the parameters that influence the detection of $B$-modes, and then decide for the type of experiment 
we have to build next.

\bibliography{cosmo,cosmo_preprints}

\end{document}